# Linear Dichroism Conversion in Quasi One-Dimensional Perovskite Chalcogenide

Jiangbin Wu[1], Xin Cong[2,3], Shanyuan Niu[4], Fanxin Liu[5], Huan Zhao[1], Zhonghao Du[1], Jayakanth Ravichandran[4,*], Ping-Heng Tan[2,3,*], Han Wang[1,4,*]

[1]Ming Hsieh Department of Electrical Engineering, University of Southern California, Los Angeles, CA 90089, USA

[2]State Key Laboratory of Superlattices and Microstructures, Institute of Semiconductors, Chinese Academy of Sciences, Beijing 100083, China

[3] College of Materials Science and Opto-Electronic Technology & CAS Center of Excellence in Topological Quantum Computation, University of Chinese Academy of Science, Beijing 100049, China.

[4]Mork Family Department of Chemical Engineering and Materials Science, University of Southern California, Los Angeles, CA 90089, USA.

[5]Collaborative Innovation Center for Information Technology in Biological and Medical Physics, and College of Science, Zhejiang University of Technology, Hangzhou 310023, P. R. China

*E-mail: han.wang.4@usc.edu (H.W.), phtan@semi.ac.cn (P.-H.T.), jayakanr@usc.edu (J.R.)

**Abstract –** Anisotropic photonic materials with linear dichroism property are crucial components in many sensing, imaging and emerging light based quantum communication applications. Such materials play an important role as polarizers, filters, wave-plates, phase-matching elements in photonic devices and circuits. Conventional crystalline materials with optical anisotropy typically show strong unidirectional linear dichroism over a broad wavelength range and even bandgap variations along different crystal directions. On the other hand, the linear dichroism conversion phenomenon - the orthogonal switch in the principle dichroism axis of the material across different optical wavelength – has not been observed in crystalline materials. Here, we report the investigation of the unique linear dichroism conversion phenomenon in quasi-one-dimensional (quasi-1D) hexagonal perovskite chalcogenide $BaTiS_3$. The material shows record level of optical anisotropy within the visible wavelength range. In contrast to conventional anisotropic optical materials, the linear dichroism polarity in $BaTiS_3$ makes an orthogonal change at an optical wavelength corresponding to the photon energy of 1.78 eV. First principle calculations reveal that this anomalous linear dichroism conversion behavior originates from different selection rules of the optical transitions from the parallel bands in the $BaTiS_3$ material. Wavelength dependent polarized Raman spectroscopy further confirms this phenomenon. Such material with linear dichroism conversion property can facilitate new ability to control and sense the energy, phase, and polarization of light, and lead to novel photonic devices such as polarization-wavelength selective detectors and lasers for multispectral imaging, sensing and optical communication applications.

**Introduction**

Linear dichroism refers to the difference between the absorption of light polarized along the parallel and perpendicular directions with respect to a specific orientation axis of a natural or artificial crystal.[1] Strong linear dichroism can be obtained by designing the micro/nanostructures in metamaterials, which can be used in optical components including polarizers and wave plates.[2-6] In naturally occurring crystals, the linear dichroism typically arises from the reduced symmetry in the crystal lattice that depends on the specific crystal structure and the elemental composition.[7-14] Besides improving the performance of traditional optical components, these anisotropic crystals can enable polarization-sensitive photonic devices[15-20] important for many emerging sensing and communication applications.

Materials with strong optical anisotropy usually have a reduced level of crystal symmetry which often exists among low-dimensional crystals, such as two-dimensional (2D)[10] and one-dimensional (1D) materials[21], as well as in bulk materials[22]. Layered black phosphorus (BP) with reduced in-plane crystal symmetry has attracted considerable attention due to the strong anisotropy in its vibrational[23], optical[9] and electrical properties[10]. Other 2D and 1D materials, such as $ReX_2$ (X=S and Se)[12, 13], AX (A=Ge, Sn, Tl; X=S and Se)[24-28] and carbon nanotubes[29], have also demonstrated linear dichroism. However, in these materials, the linear dichroism polarity is uniform within a very broadband wavelength range, i.e. the absorption of light polarized along one particular direction is always stronger than that along other directions.[16, 24, 25] Recently, members of the ternary perovskite chalcogenides family[30, 31] including $BaTiS_3$[32] and $Sr_{1-x}TiS_3$[33] were demonstrated to possess pronounced anisotropic optical responses in the mid-wave IR (MWIR) and long-wave IR (LWIR) spectral ranges. In those materials, parallel quasi-one-dimensional (quasi-1D) chains are formed by face-sharing the $TiS_6$ octahedra. The giant birefringence and linear dichroism in the MWIR and LWIR region result from the large structural and chemical anisotropy between the intra and inter-chain directions.

In this work, we reveal a unique linear dichroism conversion phenomenon in the quasi-1D $BaTiS_3$ material. Observed using reflectance anisotropy spectroscopy, the ultra-high level of linear dichroism in $BaTiS_3$ in the visible range exceeds that in any other crystalline materials, and an anomalous linear dichroism conversion occurs at the photon energy of 1.78 eV. Density functional theory (DFT) calculations show that the linear dichroism conversion originates from the polarization-resolved localized joint density of states of the material due to its parallel band structure in this optical transition energy range. Furthermore, wavelength dependent polarized Raman spectroscopy also evidenced this linear dichroism phenomenon. The resulting unique wavelength and crystal-orientation dependent optical properties of BaTiS$_3$ can open door to new opportunities for enabling wavelength tunable and polarization sensitive photonic devices desired for many applications in optical communication, imaging and sensing.

**Results**

Large single-crystal platelets of BaTiS$_3$ with lateral dimensions on the order of several millimeters were grown by the vapor transport method using iodine as a transport agent.[32] The quasi-1D BaTiS$_3$ crystallizes in a hexagonal structure as shown in Figure 1a and 1b. In BaTiS$_3$, BaS$_6$ octahedra sharing common faces are connected to form parallel chains along the *c* axis, which is the axis of six-fold rotation symmetry in the material. BaTiS$_3$ belongs to an emerging family of inorganic perovskite materials with a general chemical formula, ABX$_3$, where A is typically an alkaline earth or alkali metal ion, and B is a transition metal ion surrounded by six anions (X).[30-33]. The A ions are encapsulated within the hexagonal BX$_6$ structure. The interaction within the BX$_6$ cell is much stronger than that among the BX$_6$ cells. This is analogous to two-dimensional layered materials, which has strong covalent bonds within the lattice plane but only weak van der Waals coupling between the layers. Thus, bulk BaTiS$_3$ can be exfoliated into thin flakes with thickness below tens of nanometers, as shown in the optical micrograph in Figure 1c. The atomic force microscope (AFM) is used to confirm the thickness of the flake (Figure S1). Energy-dispersive X-ray spectroscopy (EDS) mapping shows the expected elemental composition (Figure 1d). High resolution transmission electron microscopy (HR-TEM) images of the crystals along the *c*-axis (Figure 1f) and *a*-axis (Figure 1g) clearly reveal the presence of the hexagonal arrangement of the chains and parallel 1D chains along the *c* axis. Selected-area electron diffraction patterns are also available in the illustrations of Figures 1e and 1f, which confirm the hexagonal and square section of *c*- and *a*-axis, respectively.

To experimentally characterize the linear dichroism of BaTiS$_3$, a supercontinuum laser source (NTK SuperK EXTREME OCT) is employed, whose laser beam is polarized by a broadband polarizer (see Method). A broadband half-wave plate is used to tune the polarization angle of the incident laser beam, as shown in Figure 2a. Here, we define the X axis of the incident light as being parallel to the *a*-axis of the BaTiS$_3$ crystal while the Y axis of the incident laser is parallel to the *c*-axis. The reflected light is collected by a charge-coupled device (CCD) camera. All the reflectance measurements are normalized with respect to a reference spectrum measured from a silver mirror. The measured reflectance with the incident laser polarization along both the X- and Y-directions is presented in Figure S2. From a Kramers-Kronig transform[34] of the reflectance data we calculate the absorption coefficient as a function of frequency (see supplementary information for details).

Figure 2b shows the absorption spectra with the incident laser polarization along both the X- and Y-directions ($A_x$ and $A_y$) in a photon energy range from 1.5 eV to 2.7 eV. Among other existing materials showing linear dichroism, it is commonly observed that the absorption of light polarized along one crystal direction is always stronger than that along other crystal directions over a broad wavelength range. However, the linear dichroism polarity in BaTiS$_3$ shows a significant dependence on wavelength. As shown in Figure 2b, $A_y$ has much stronger intensity than $A_x$ in the energy range from 1.6 to 1.78 eV. As the photon energy increases above 1.78 eV, however, the dominant axis of the linear dichroism changes to a different direction that is almost orthogonal to that below 1.78 eV. When the photon energy reaches above 2.15 eV, $A_x$ becomes close to $A_y$ and the linear dichroism is much weaker. Hence, $A_x$ and $A_y$ show a clear crossover at 1.78 eV, i.e. the linear dichroism conversion at 1.78 eV. Here we define the degree of linear polarization (DLP) as $(A_y - A_x)/(A_x + A_y)$, which has values within the range from -1 to 1. A DLP value between -1 and 0 means $A_y<A_x$, and a DLP value between 0 and 1 will indicate $A_y>A_x$. Moreover, the larger magnitude of DLP indicates stronger linear dichroism. When the photon energy is below 1.78 eV, the DLP is observed to be less than 0, and it becomes greater than 0 for photon energies above 1.78 eV. At the photon energy range between 1.65 eV and 1.71 eV, the magnitude of the DLP reaches above 0.9, indicating the ultra-high level of linear dichroism in BaTiS$_3$ within this wavelength range. This DLP is much higher than that in other naturally existing materials with strong linear optical dichroism including black phosphorus[9] and ReS$_2$[35]. It is also close to the record value achieved in the artificial plasmonic metasurfaces[2].

Anisotropic absorption spectra are observed by changing the polarization angle (θ) (it's defined as the angle between the polarization of incident beam and *a*-axis) from 0° to 360°, as shown in Figure 2c. At the vicinity of 1.7 eV photon energy, the maximum reflectance is observed at the angle of 90° and 270°, and for photon energy around 2.0 eV, the maximum reflectance is observed at the angle of 0° and 180°. Figure 2d shows the polar plots of the reflectance at the photon energies of 1.70 eV and 1.96 eV, respectively, which is consistent with the results shown in Figure 2b and 2c. As the polarization angle $\theta$ is varied, the absorption (*A*) shows periodic changes, which can be fitted by: $A = (A_{max} - A_{min})\cos2(\theta - \theta_0) + A_{min}$, where $\theta_0$ denotes the reference polarization angle at which the absorption coefficient reaches the maximum. Thus, under the excitation energy of 1.70 eV, the maximum reflectance occurs along the *a*-axis of the BaTiS$_3$ crystal, while under the excitation energy of 1.96 eV, the maximum reflectance is reached along the *c*-axis of the crystal. The results in Figure 2c hence reveals the linear dichroism conversion in the material.

To explore the origin of the linear dichroism conversion, the calculation of the polarization-resolved absorption spectra of BaTiS$_3$ is performed using DFT (see Method), as shown in Figure 3a. One prominent peak is observed along the *a*-axis at 2.0 eV excitation photon energy. On the other hand, along the *c*-axis, there are three narrow peaks at 1.15 eV, 1.6 eV and 2.33 eV, resulting in a valley at 1.9 eV. Hence, the calculated polarization-resolved absorption spectra predict a linear dichroism conversion from 1.5 to 2.3 eV and the crossover occurs at 1.8 eV. The calculation results agree well with the experimental data. When the photon energy is larger than 2.3 eV, the theoretical results indicate that the linear dichroism becomes weak, which also agrees with the experimental results in Figure 2b. The calculated polarization-resolved reflection spectra are also shown in Figure S3, which shows good agreement with experimental reflection spectra in Figure S2. To further understand the origin of the peaks in the absorption and reflection spectra, the band structure and density of states (DOS) in BaTiS$_3$ are calculated as shown in Figure 3b. Optical transition matrix element (dipole-transition selection rule) is also considered in the calculation. The optically-allowed transitions between a series of parallel bands corresponding to the narrow reflection peaks are indicated in Figure 3b.

Along the *a*-axis, the peak at 2.0 eV is assigned to the transition between the $p_z$ orbital of the S atom to the $d_{yz}$ and $d_{xz}$ orbitals of the Ti atom. Along the *c*-axis, the peak at 1.15 eV is due to the transition between the $p_z$ orbital of the S atom to the $d_z^2$ orbital of the Ti atom. The transition between the $p_x$ and $p_y$ orbitals of the S atom and the $d_{yz}$ and $d_{xz}$ orbital of the Ti atom lead to the peak at 1.6 eV. Moreover, the peak at 2.3 eV originates from the transition between the $p_z$ orbital of the S atom and the $d_{xy}$ and $d_{x^2-y^2}$ orbitals of the Ti atom. The orbitals corresponding to the relevant energy bands are labeled in Figure 3b. Figure 3c-3g show the partial charge densities for the states in Figure 3b. It is worth noting that the highlighted conduction and valence bands in Figure 3b are almost parallel to each other, which resembles the band structure features in twisted bilayer graphene.[36-38] These parallel bands would lead to the localized joint density of states (absorption peaks) at the specific photon energies. In the twisted bilayer graphene, the localized joint density of states leads to absorption wavelength (photon energy) that is highly dependent on the twisting angles between the two graphene layers. Here, the wavelength dependence of the linear dichroism polarity (linear dichroism conversion) arises in BaTiS$_3$ due to the absorption peaks along the two principal crystal directions being strongly localized at different energies.

The Raman spectroscopy can reveal polarization-resolved phonon characteristics in many nanomaterials. Indeed, the Raman tensor of low-symmetry crystal always shows a high degree of anisotropy.[39-41] However, the Raman scattering process typically involves both electron-phonon (Raman tensor) and electron-photon interactions (dipole-transition selection rule).[40] According to the second-order perturbation theory, the polarization-resolved Raman intensity $I(\theta) \propto |M_{eR}^i(\theta) \cdot M_{ph}(\theta) \cdot M_{eR}^s(\theta)|^2$, where $M_{eR}^i(\theta)$ and $M_{eR}^s(\theta)$ are the optical transition matrix elements of the incident and scattered light, respectively, and $M_{ph}(\theta)$ is the matrix element of the electron–phonon interaction.[42] The Raman spectra of BaTiS$_3$ are measured under the excitation of 1.71 eV, 1.96 eV and 2.33 eV lasers with the incident polarization parallel to the *a*-axis, as shown in Figure 4a. There are two prominent peaks at 186 cm$^{-1}$ and 380 cm$^{-1}$. The normal mode displacements of these two phonon modes are calculated using density functional perturbation theory (DFPT) and shown in Figure 4b. Because the BaTiS$_3$ has a D$_{6h}$ symmetry, according the vibrational displacement, we classified these two phonon modes as $A_{1g}$, and denoted them as $A_{1g}^1$ and $A_{1g}^2$ based on their frequencies, respectively.

Since the $A_{1g}^1$ mode disappears at 1.71 and 1.96 eV excitations, here we focus on discussing the $A_{1g}^2$ mode. Figure 4c and 4d show the polarization-resolved Raman spectra and intensity of the $A_{1g}^2$ mode under the three excitation wavelengths at 1.71 eV, 1.96 eV and 2.33 eV. The principal polarization orientation (the orientation of the strongest intensity) of the Raman spectra between 1.96 eV and 2.33 eV are almost perpendicular (~85°), and those of 1.71 eV and 2.33 eV are parallel to each other. To understand the underlying mechanism, we consider both electron-phonon and electron-photon interactions. The Raman tensor of $A_{1g}$ mode is[43]:

$$A_{1g} = \begin{bmatrix} a & 0 & 0 \\ 0 & a & 0 \\ 0 & 0 & b \end{bmatrix}.$$

For the 2.33 eV excitation where there is almost no linear dichroism in BaTiS$_3$, as shown by both experiment and theory in Figure 2b and 3a, the anisotropy in $M_{eR}^i(\theta)$ and $M_{eR}^s(\theta)$ can be neglected. Thus, $I(\theta) \propto |M_{ph}(\theta)|^2 \propto |a\sin\theta|^2 + |b\cos\theta|^2$ according to Raman tensor[42]. Therefore, the principal polarization orientation under the excitation of 2.33 eV is determined by the ratio of $a^2/b^2$. By fitting the experimental data, we can extract the value of $a^2/b^2$ to be 1.62. This indicates the dependence of the phonon mode intensity ($|M_{ph}(\theta)|^2 \propto 0.62|\sin\theta|^2 + 1$) on the polarization direction and the strong electron-phonon interaction occurs along the crystal direction that is almost parallel to the $c$ axis, as shown in Figure 4d. At other wavelengths (for example, 2.71 eV, as shown in Figure S4) where the linear dichroism is also week (as shown in Figure S2 and S3), the linear dichroism also can be neglected, the principal polarization orientation of the $A_{1g}^2$ mode should be the same as that under the excitation of 2.33 eV and with the similar $a^2/b^2$ (as shown in Figure S4). However, due to the presence of strong linear dichroism at 1.96 eV excitation energy, the principal polarization orientation of the $A_{1g}^2$ mode becomes almost parallel to that of the absorption, *i.e.* almost perpendicular (~85°) to that of the $A_{1g}^2$ mode under the excitation of 2.33 eV photon energy. It indicates that the linear dichroism plays a more dominant role at this excitation wavelength. As a result, the principal polarization orientation of the $A_{1g}^2$ mode observed in Raman spectroscopy is dominated by the absorption instead of the electron-phonon-interaction, which further confirms the strong linear dichroism of the BaTiS$_3$ under the excitation of 1.96 eV. Under the 1.71 eV excitation energy where it has the orthogonal linear dichroism compared with 1.96 eV excitation, the principal polarization orientation of the $A_{1g}^2$ mode in the Raman spectrum is parallel to that with 2.33 eV excitation and shows even higher level of anisotropy. As shown in Figure 4c and 4d, the ratio between the maximum and minimum of θ-dependent Raman intensities: I(θ)$_{max}$/I(θ)$_{min}$ is 1.62 under the excitation of 2.33 eV, and I(θ)$_{max}$/I(θ)$_{min}$ is 3.44 for 1.71 eV excitation. Since the polarities of both the absorption and the electron-phonon-interaction align with each other (along the *c*-axis) at 1.71 eV excitation, the level of anisotropy in the Raman intensities becomes significantly enhanced compared to that with the 2.33 eV excitation. Therefore, the wavelength dependent polarities of the Raman intensities further confirm the strong linear dichroism and linear dichroism conversion properties in the BaTiS$_3$ crystal.

**Conclusion**

In summary, the strong and anomalous linear dichroism in the *a-c* plane of quasi-1D BaTiS$_3$ is observed through the reflectance anisotropy spectroscopy. The DLP in the material reaches a record peak value of 0.9 within the visible wavelength range. The linear dichroism reverses its polarity at the 1.78 eV excitation energy due to the localized joint density of states along both the *a*- and *c*- directions of the crystal resulting from the parallel band structure in BaTiS$_3$. The Raman spectroscopy measurements under the 1.71 eV, 1.96 eV and 2.33 eV excitations further confirm the strong linear dichroism and linear dichroism conversion properties in the material. The unique linear dichroism conversion property found in BaTiS$_3$ ultrathin flakes can lead to novel photonic detection device, which can provide dynamically tunable responses with respect to both the wavelength and the polarization of the incident light, with broad applications in optical sensing, communications, multi-spectral imaging.

**Method**

**SEM, TEM and EDS Measurements**: SEM and TEM images were taken by JEOL JSM 7001. 15 kv accelerating voltage was used in the imaging. Samples for TEM cross sections were prepared by focused ion beam (FIB) in the direction parallel and perpendicular to *c*-axis. Magnifications from ×100 to ×15,000 were used for the EDS imaging, and the obtained chemical composition ratios were consistent between these magnifications.

**DFT Calculations:** DFT code Vienna ab initio simulation package (VASP)[44] was used to perform the structural relaxation within the projector augmented wave method[45, 46] and a plane-wave basis. The exchange correlation potential is treated within the generalized gradient approximation (GGA). The electronic structures, charge density and optical properties were then obtained using the GGA + U method[47]. U is the Coulomb parameter. A 16×16×16 k-mesh is used to sample the Brillouin zone for $BaTiS_3$ unit cell. The energy cutoff for the plane-wave basis is 400 eV. All atoms are fully relaxed until the residual force per atom is smaller than $1\times10^{-5}$ eV•Å$^{-1}$. Phonon frequencies are calculated using DFPT[48], as implemented in VASP.

**Raman measurement:** Raman spectra are measured in back-scattering at room temperature with a Jobin-Yvon HR800 Raman system, equipped with a liquid nitrogen-cooled CCD, a ×100 objective lens (numerical aperture (NA)=0.90) and several gratings. The excitation energies are 2.33 eV from a solid laser, 1.96 eV from a He-Ne laser and 1.71 eV from a Ti:Saphire laser. The resolution of the Raman system at 2.33 eV is 0.49 cm$^{-1}$ per CCD pixel. The typical laser power is ~500 μW, to avoid sample heating.

**Figures**

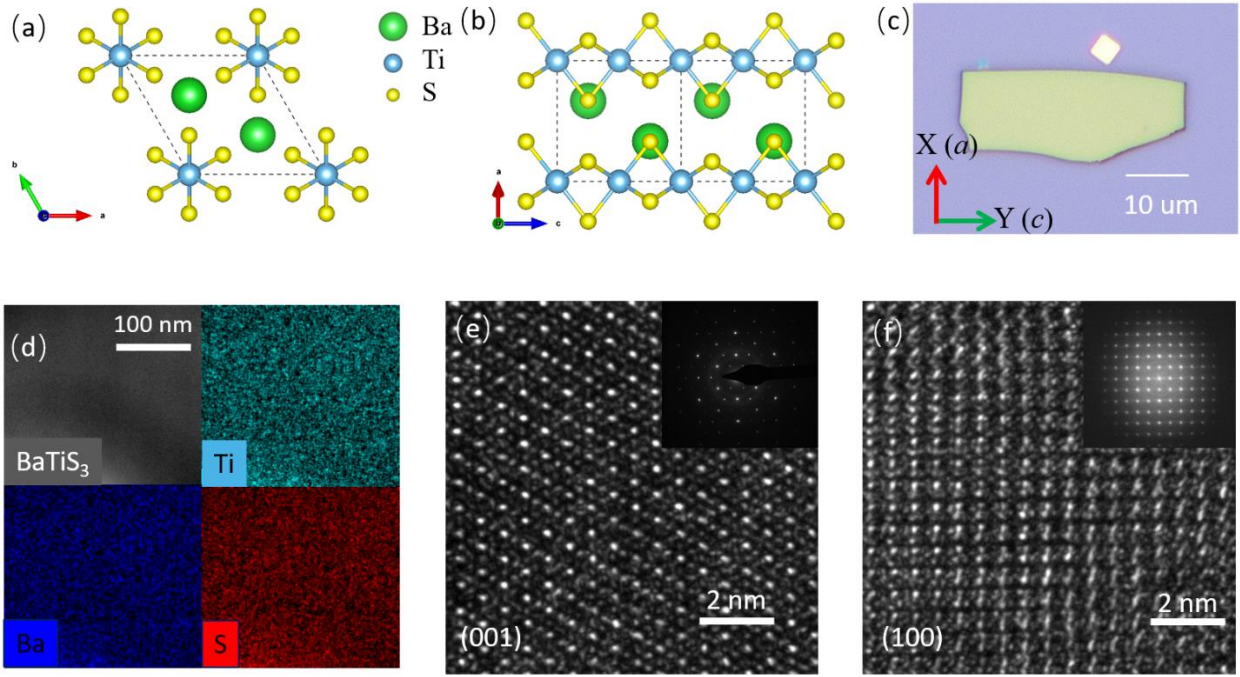

Figure 1. (a) BaTiS$_3$ crystal structure viewed along the *c*-axis, showing hexagonal symmetry. (b) BaTiS$_3$ crystal structure viewed along the *a*-axis, showing the TiS$_6$ chains parallel to the *c*-axis. (c) Optical micrograph of BaTiS$_3$ sample mechanically cleaved along *a-c* plane. (d) EDS mapping of the barium (blue), titanium (green) and sulfur (red) elements in the a BaTiS$_3$ crystal (top left). The scale bar is 100 nm. Atomic resolution TEM images of BaTiS$_3$ viewed along the *c*-axis (e) and *a*-axis (f). The insets are the corresponding electron diffraction patterns of the TEM images.

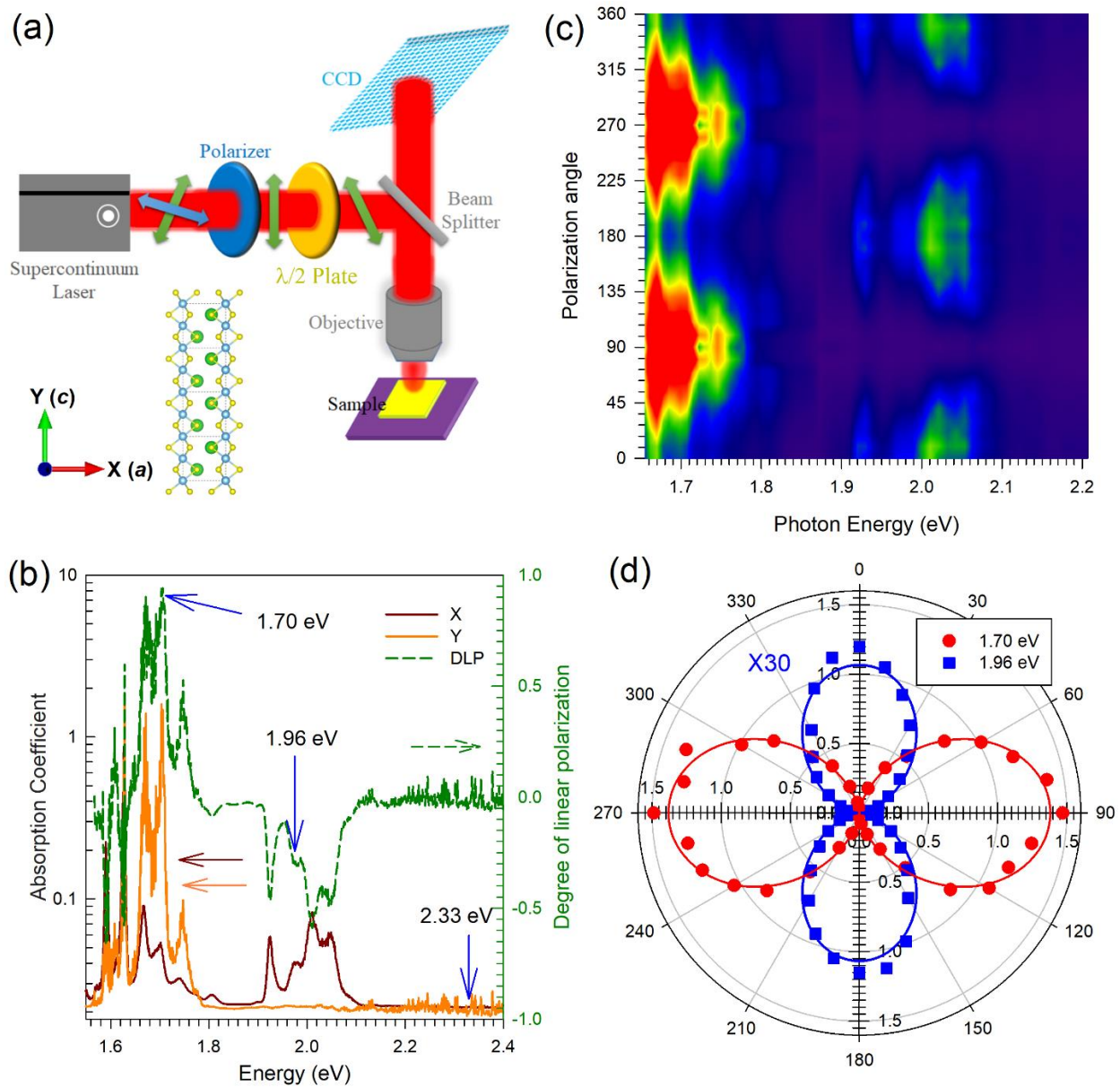

Figure 2. (a) Schematic of the experimental setup for measuring the polarization conversion. (b) The measured absorption coefficient for the X- and Y-polarized light and degree of linear polarization (DLP) in the visible range. (c) The absorption spectra as a function of both the polarization angle and the excitation photon energy. The color pattern in the spectra indicates the anisotropy in the reflection coefficients. (d) Polar plots of the absorption coefficient as a function of the polarization angle of the incident laser at the excitation photon energy of 1.70 eV and 1.96 eV. The red dots and blue squares are the experimental data. The red and blue lines are the fitted curves.

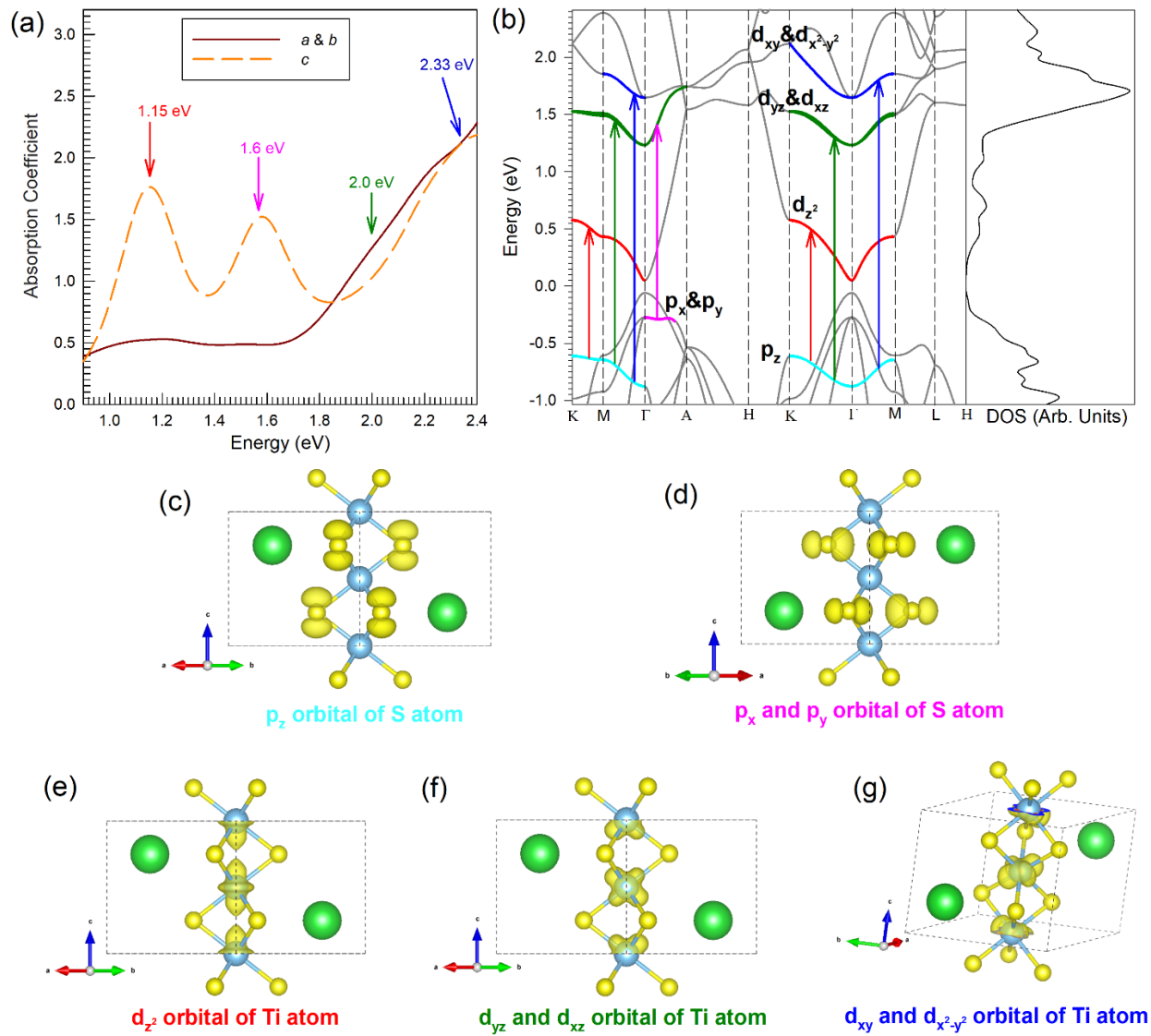

Figure 3. (a) The calculated absorption coefficient spectra for excitation light polarized parallel (orange) and perpendicular (brown) to the $c$-axis. (b) The calculated electronic band structure and density of states (DOS) of $BaTiS_3$. The orbitals of the corresponding bands are highlighted and marked. The calculated partial charge densities of the orbitals are shown in (c) $p_z$ in S atom, (d) $p_x$ and $p_y$ in S atom, (e) $d_{z^2}$ in Ti atom, (f) $d_{yz}$ and $d_{xz}$ in Ti atom, and (g) $d_{xy}$ and $d_{x^2-y^2}$ in Ti atom.

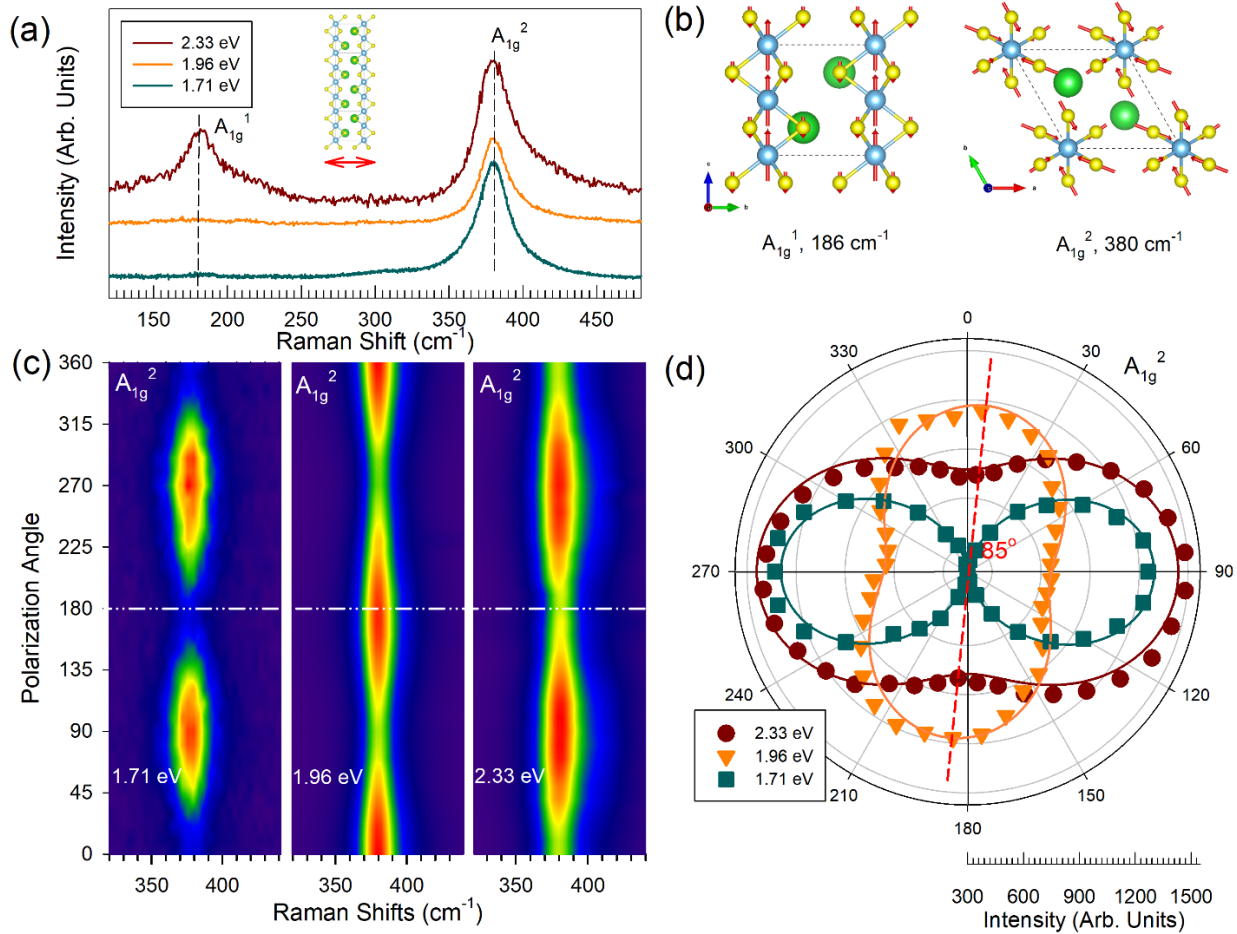

Figure 4. (a) Raman spectra of BaTiS$_3$ measured using 2.33 eV, 1.96 eV and 1.71 eV excitation lasers with the laser polarized perpendicular to the *c*-axis of the crystal. (b) Displacement of the corresponding phonon modes in (a) calculated by DFPT. (c) Anisotropic Raman spectra in the region of $A_{1g}^2$ mode plotted as a function of the polarization angle and the Raman shift for the three excitation conditions at 1.71 eV, 1.96 eV, and 2.33 eV, respectively. (d) Polar plots of the $A_{1g}^2$ mode intensity as a function the polarization angle at 2.33 eV, 1.96 eV and 1.71 eV excitations. The brown dots, orange triangles and cyan squares are the experimental data. The brown, orange and cyan lines are the fitted curves.

# Supporting Information

# Linear Dichroism Conversion in Quasi One-Dimensional Perovskite Chalcogenide

Jiangbin Wu[1], Xin Cong[2,3], Shanyuan Niu[4], Fanxin Liu[5], Huan Zhao[1], Zhonghao Du[1], Jayakanth Ravichandran[1,4,*], Ping-Heng Tan[2,3,*], Han Wang[1,4,*]

[1]Ming Hsieh Department of Electrical Engineering, University of Southern California, Los Angeles, CA 90089, USA

[2]State Key Laboratory of Superlattices and Microstructures, Institute of Semiconductors, Chinese Academy of Sciences, Beijing 100083, China

[3] College of Materials Science and Opto-Electronic Technology & CAS Center of Excellence in Topological Quantum Computation, University of Chinese Academy of Science, Beijing 100049, China.

[4]Mork Family Department of Chemical Engineering and Materials Science, University of Southern California, Los Angeles, CA 90089, USA.

[5] Collaborative Innovation Center for Information Technology in Biological and Medical Physics, and College of Science, Zhejiang University of Technology, Hangzhou 310023, P. R. China

*E-mail: han.wang.4@usc.edu (H.W.), phtan@semi.ac.cn (P.-H.T.), jayakanr@usc.edu (J.R.)

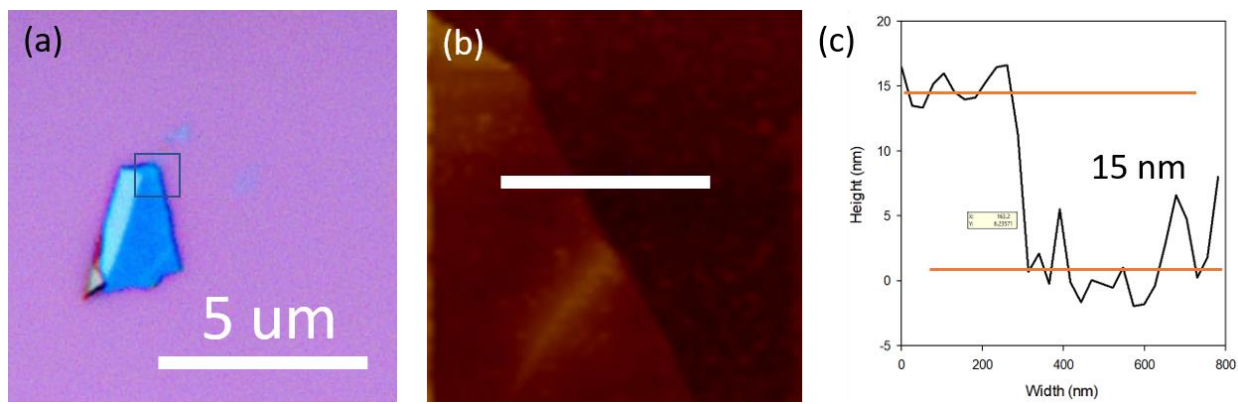

Figure S1: (a) Optical image of BaTiS$_3$ exfoliated onto Si/SiO$_2$ substrate. b) Atomic force microscopy image of BaTiS$_3$ flake marked by the blue square in (a). (c) Height data with respect to the distance along white line marked in (b).

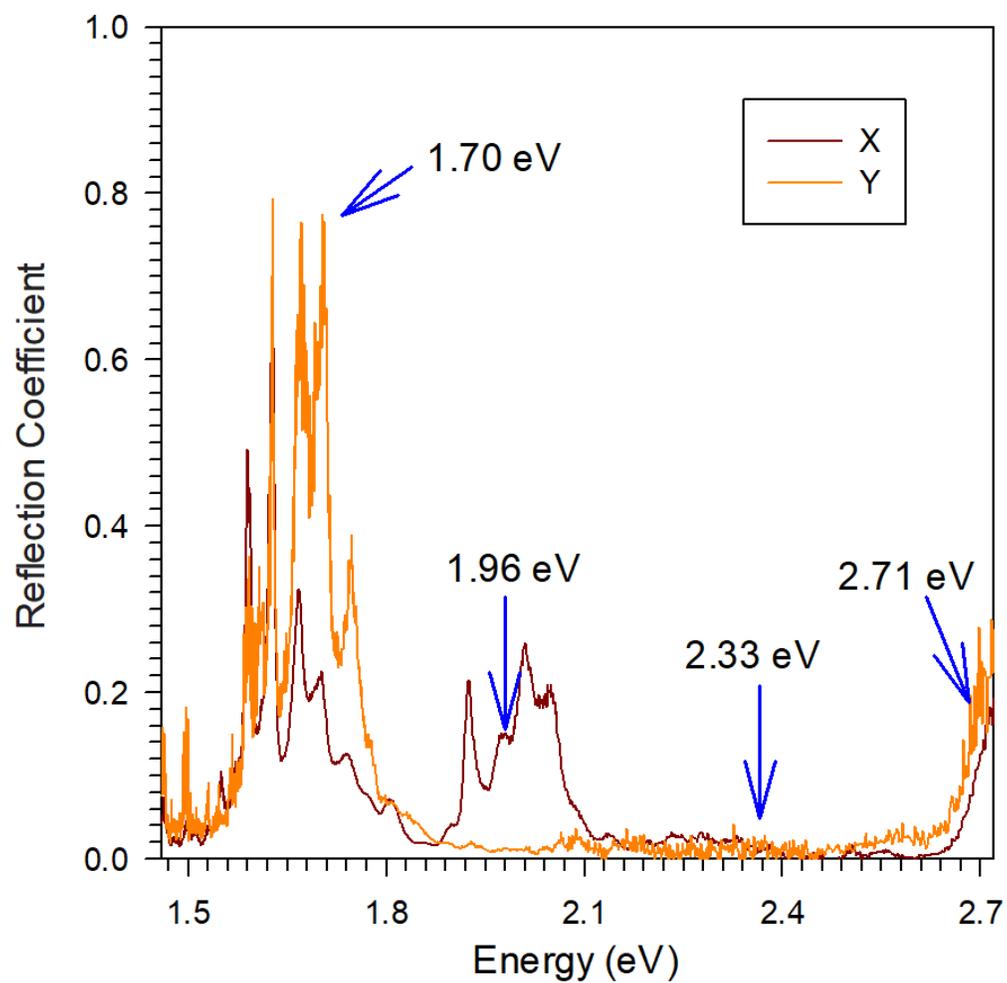

Figure S2 The measured reflectance for the X- and Y-polarized light in the visible range.

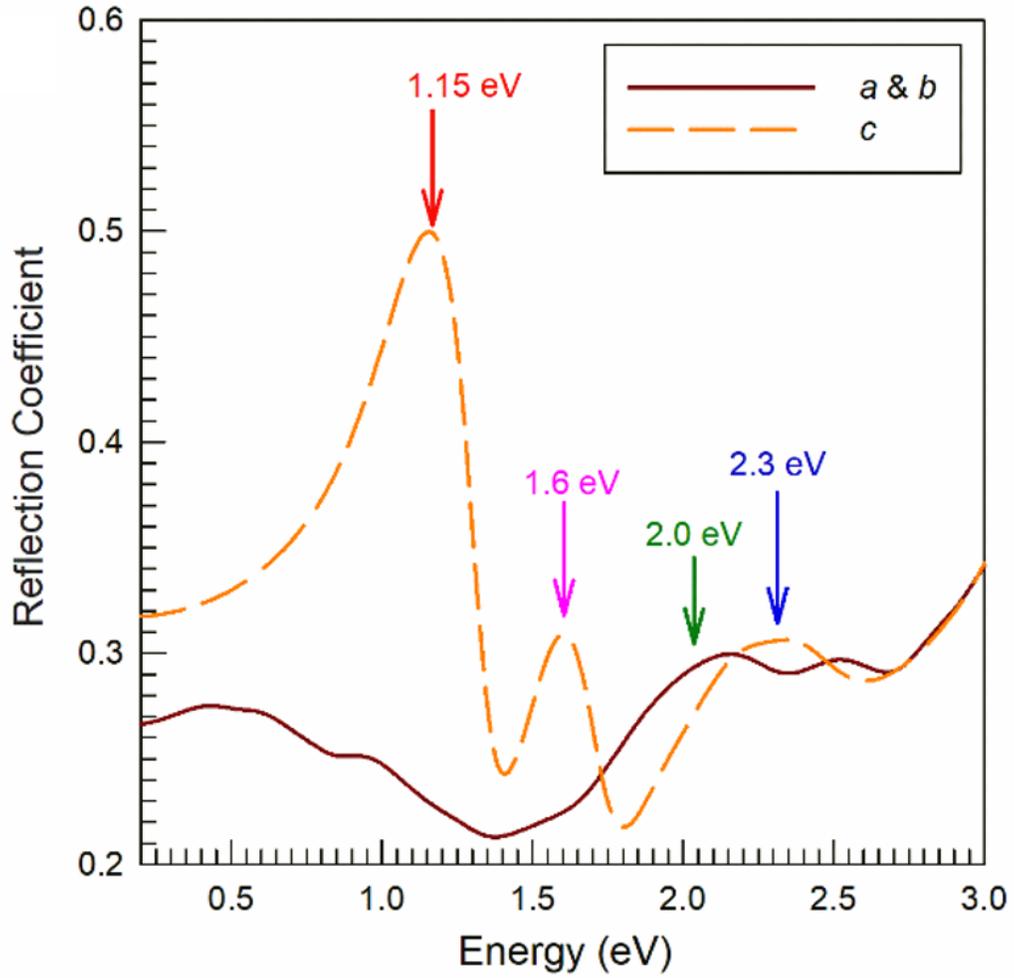

Figure S3 The calculated reflection coefficient spectra for excitation light polarized parallel (yellow) and perpendicular (brown) to the c-axis.

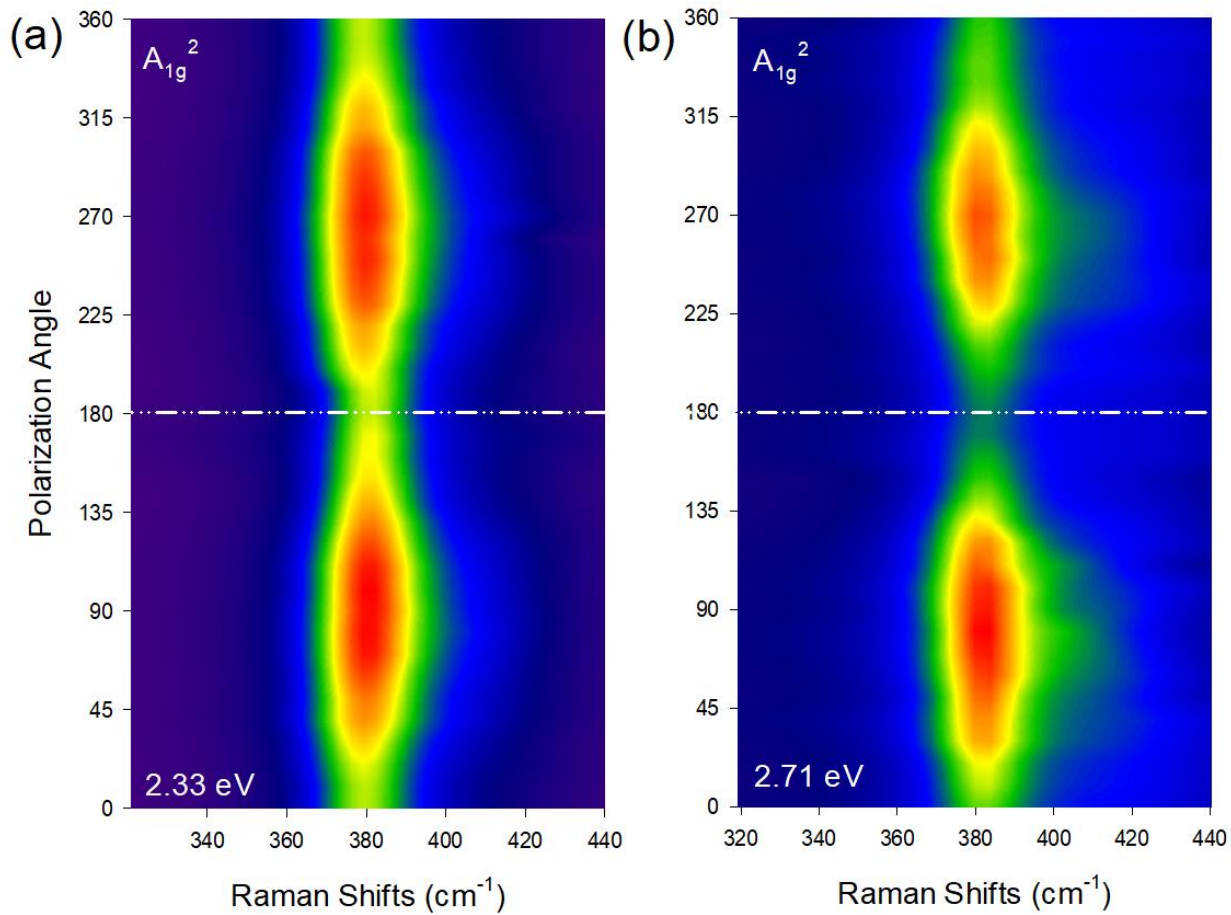

Figure S4: 2D colormap of anisotropic Raman spectra in the region of $A_{1g}^2$ mode under the excitation of 2.33 eV (a) and 2.71 eV (b).

## Note 1: Kramers-Kronig Analysis

From a Kramers-Kronig transform of the reflectance data we calculate the index of refraction $n(\omega)$ and the index of the extinction coefficient $\kappa(\omega)$ as functions of frequency. These properties are fundamental material characteristics, as they effectively define its optical response to light.

The Kramers-Kronig relations are based on the requirement of causality and the fact that the real and imaginary parts of a response function are always related through a dispersion relation.[1-3] For reflectance measurements the response function $r(\omega)$ is the product of an amplitude and a phase:

$$r(\omega) = \frac{E_{refl}}{E_{incid}} = \rho(\omega)\exp[i\Theta(\omega)],$$

where $E$ is the incident or reflected electric-field vector. The measured reflectance is

$$R(\omega) = \left|\frac{E_{refl}}{E_{incid}}\right|^2 = \rho(\omega)^2$$

and the Kramers-Kronig integral relates the phase shift to the reflectance:

$$\Theta(\omega) = \frac{\omega}{\pi}\int_0^\infty \frac{\ln[R(\omega')] - \ln[R(\omega)]}{\omega^2 - \omega'^2} d\omega'.$$

Using

$$r(\omega) = \frac{(n-1)+i\kappa}{(n+1)+i\kappa},$$

one can obtain the real and imaginary parts of the complex refractive index. The refractive index $n$ is

$$n(\omega) = \frac{1-R(\omega)}{1+R(\omega)-2[R(\omega)]^{1/2}\cos[\Theta(\omega)]},$$

and the extinction coefficient $\kappa$ is

$$\kappa(\omega) = \frac{2[R(\omega)]^{1/2}\sin[\Theta(\omega)]}{1+R(\omega)-2[R(\omega)]^{1/2}\cos[\Theta(\omega)]}.$$

The absorption coefficient a can be obtained as

$$\alpha(\omega) = 4\pi\kappa\omega,$$

where ω is the frequency in wave numbers (cm$^{-1}$). Thus, by a Kramers-Kronig analysis of reflectance data, we can obtain the optical constants of a material.